\documentclass[aip,jcp,amssymb,amsmath,reprint,floatfix]{revtex4-2}

\usepackage[utf8]{inputenc}
\usepackage[T1]{fontenc}
\usepackage{braket}
\usepackage{graphicx}
\usepackage[english]{babel}
\newcommand{\safeincludegraphics}[2][]{%
  \IfFileExists{#2}{\includegraphics[#1]{#2}}{%
    \fbox{\parbox{\linewidth}{\centering Missing file: \texttt{\detokenize{#2}}}}%
  }%
}

\usepackage{dcolumn}% Align table columns on decimal point
\usepackage{etoolbox}
\usepackage{physics}
\usepackage{hyperref}
\usepackage{amsmath,amssymb}
\usepackage{here}
\usepackage{xcolor}
\usepackage{mathrsfs}
\newif\ifincludeTCLfigure
\includeTCLfiguretrue 

\begin{document}
% ====================
% TITLE & AUTHORS
% ====================
\title{
Reduced Dynamical Maps in Finite Temperature Vibronic Coupling Models via Choi Matrices: Numerical Methods and Applications}

\author{Raffaele Borrelli}
\affiliation{DISAFA, University of Torino, Torino, Italy}
\email{raffaele.borrelli@unito.it}
\author{Hideaki Takahashi} 
\affiliation{DISAFA, University of Torino, Torino, Italy}
\email{hideaki.takahashi@unito.it}
% ====================
% ABSTRACT
% ====================
\begin{abstract}
We present a streamlined implementation of a computational framework for constructing and
analyzing reduced dynamical maps for complex system--bath models at finite
temperature. The methodology is based on three established ingredients of quantum dynamics: 
the Choi--Jamio{\l}kowski isomorphism
for the representation of quantum channels, thermofield (TFD) purification of thermal environments, and tensor-train (TT) propagation of the resulting enlarged pure state.
The reduced map is obtained from a single unitary propagation in a
thermofield-doubled Hilbert space and represented in matrix form through
the Choi--Jamio{\l}kowski isomorphism.  The TFD evolution is implemented in the
TT representation, enabling efficient propagation of
high-dimensional purified thermal states. 
We illustrate the methodology for exciton transfer in the
Fenna--Matthews--Olson complex with site-dependent structured spectral
densities represented by discretized bosonic environments.
The resulting maps are used to analyze decoherence, relaxation, and finite-memory effects,
and to assess the crossover to an effectively time-local
description.  The proposed approach provides a route to compute reduced
propagators and to post-process them into memory kernels, transfer tensors, and
effective kinetic rate descriptions for complex molecular systems.
\end{abstract}

\maketitle

% ====================
% INTRODUCTION
% ====================
\section{Introduction}
\label{sec:intro}

Open quantum dynamics plays a central role in chemical physics, where energy and
charge transport, vibronic relaxation, and nonlinear spectroscopic signals are
shaped by the interplay between coherent evolution and environmental
fluctuations.  In realistic molecular and condensed-phase settings the system is
coupled to a structured, finite-temperature environment whose correlation time
can be comparable to intrinsic system time scales, giving rise to dissipation,
decoherence, and memory effects.

A wide range of theoretical tools has been developed to treat such problems.  
Perturbative or Markovian master equations (\textit{e.g.} Redfield- and
GKLS-type approaches)
\cite{Redfield1965AiMaOR,Lindblad1976CMP}
are computationally efficient, but their accuracy is restricted by assumptions
on coupling strength, bath memory time, and separation of time scales.  
Numerically accurate methods that remain reliable for strong coupling and
structured environments---including hierarchical equations of motion
(HEOM),\cite{Tanimura1990PRA,TanimuraKubo1989JPSJ,Tanimura2020JCP} path-integral
approaches (QUAPI, TEMPO, and related
schemes)\cite{Makri1991CPC,Makri2018TJoCP,StrathearnEtAl2018NC}, and
wavefunction-based tensor-network methods such as (ML-)MCTDH and
(time-dependent) DMRG---have therefore become essential for quantitative
simulations.\cite{BeckEtAl2000PR,BurghardtEtAl2008JCP,Wang2015JPCA,WhiteFeiguin2004PRL}

Most workflows propagate a specific initial system state (or density matrix) and
then evaluate a limited set of observables, often populations in a chosen basis or
a specific correlation function.
While this is natural for many applications, it can hide information that is
implicit in the full reduced dynamics---for example, how different initial
coherences decay, how memory effects build up, and when the dynamics crosses
over to an effectively time-local description.  Moreover, repeating an expensive
simulation for many initial conditions becomes quickly prohibitive as the system
dimension grows.

A compact object that contains the full reduced dynamics for all factorized
initial conditions is the \emph{reduced dynamical map}
$\Phi(t,t_0)$,\cite{BreuerPetruccione2010} defined by 
\begin{equation}
\rho_S(t)=\Phi(t,t_0)[\rho_S(t_0)]
\end{equation}
where $\rho_S(t)$ is the reduced system density matrix,
with the environment traced out. 
Once $\Phi(t,t_0)$ is available, populations and coherences for arbitrary initial
$\rho_S(t_0)$ can be obtained by inexpensive post-processing, and the map can be
analyzed to obtain a wealth of information on the system-bath
entanglement dynamics.  The difficulty is that a direct construction of
$\Phi(t,t_0)$ by state-by-state propagation requires $\mathcal{O}(d_S^2)$ distinct
simulations for a $d_S$-dimensional
system.\cite{Montoya-CastilloReichman2016JCP,CerrilloCao2014PRL,LyuEtAl2023JCTC}

In this paper we exploit the \textit{well-established} Choi--Jamio{\l}kowski isomorphism to represent
$\Phi(t,t_0)$ as a matrix (the Choi matrix) and to reconstruct the full reduced map
from \emph{one} unitary propagation in an enlarged Hilbert space that includes a
replica (ancilla) of the system.\cite{Choi1975LAaiA,Jamiolkowski1972RoMP}
Our work 
builds on and adapts the strategy
recently developed by Strachan \textit{et al.}, which has been applied to fermionic reservoirs,\cite{StrachanEtAl2024TJoCP,PurkayasthaEtAl2021PRB} and recently to spin-boson problems.\cite{StrachanEtAl2025}
Here, we focus on finite-temperature linear vibronic models relevant to molecular
aggregates and condensed-phase chemical dynamics. 
The methodology relies on the Choi isomorphism
combined with Thermofield Dynamics (TFD) purification and Tensor-Train (TT) representation of many-dimensional vibronic wave functions.
While these ideas well documented in the literature, to the best of our knowledge their application to realistic linear vibronic models at finite temperature is still missing.

Specifically, TFD represents the thermal bath density matrix as a pure state in
a doubled Hilbert space, so that the full
system--environment evolution can be treated by wavefunction-based
methods.\cite{BorrelliGelin2021WCMS,BorrelliGelin2016JCP,BorrelliGelin2017SR,BlasiakEtAl2026JCP,BlasiakEtAl2025JCP,ChenZhao2017JCP,Zhao2023JCP,deVegaBanuls2015PRA}  
We then extract the Choi matrix by performing partial
traces over the (doubled) bath degrees of freedom directly in the TT
representation.  This yields a practical route to compute and store $\Phi(t,t_0)$
for complex models with many environmental modes.

In the following we discuss:
(i) a transparent derivation of the TFD--Choi construction for reduced dynamical maps at finite temperature; 
(ii) the inclusion of static (ensemble) disorder at the map level using auxiliary zero-frequency modes; and 
(iii) a set of map-based diagnostics and extrapolation strategies (time-local generators, memory kernels, transfer tensors, and population-rate reductions) that are particularly helpful in chemical-physics applications.

The paper is organized as follows.  
Section~\ref{sec:choi} introduces reduced dynamical maps and their Choi
representation.  Section~\ref{sec:tfd} derives the thermofield formulation used
to compute the Choi matrix from a single propagation, and
Section~\ref{sec:disorder} summarizes how static disorder can be incorporated.  
In Section~\ref{sec:numerics} we apply the method to exciton transfer in the
Fenna--Matthews--Olson (FMO) complex and demonstrate how the resulting maps
enable a compact analysis of decoherence, relaxation, and memory effects.  
Conclusions and outlook are given in Section~\ref{sec:conclusions}.

\section{Choi representation of the reduced dynamical map}
\label{sec:choi}

Consider a quantum system $S$ with Hilbert space $\mathcal{H}_S$ coupled to an
environment $B$ with Hilbert space $\mathcal{H}_B$.  The total Hilbert space is
$\mathcal{H}_S\otimes\mathcal{H}_B$ and the unitary propagator is
$U=\exp(-iH(t-t_0))$ (we set $\hbar=1$).  
For an initially factorized state $\rho_{SB}(t_0)=\rho_S(t_0)\otimes\rho_B$, the
reduced system state at time $t$ is 
\begin{equation}
\label{eq:map}
\Phi(t,t_0)[\rho_S] = \operatorname{Tr}_B\!\big[U(\rho_S\otimes \rho_B)U^\dagger\big]
\end{equation}
which defines the reduced dynamical map (reduced propagator) $\Phi(t):\mathcal{B}(\mathcal{H}_S)\to\mathcal{B}(\mathcal{H}_S)$.  
The evolution of any initial density matrix $\rho_S(t_0)$ can be easily obtained by applying the map as
\begin{equation}
  \label{eq:mapevo}
    \rho_S(t) = \Phi(t,t_0) \rho_S(t_0)
\end{equation}
and expectation values follow as
\begin{equation}
    \langle A(t) \rangle = \Tr{A\Phi(t,t_0)\rho_S(t_0)}.
\end{equation}

By differentiating Eq.~\eqref{eq:mapevo} with respect to time, the equation for the evolution of the reduced density matrix of the system is obtained
\begin{equation}
    \frac{d}{dt} \rho_S = [\dot{\Phi}(t,t_0)\,\Phi(t,t_0)^{-1}] \rho_S
\end{equation}
where $\mathcal{L}(t,t_0)=\dot{\Phi}(t,t_0)\,\Phi(t,t_0)^{-1}$
is the generator of the map.\cite{BreuerPetruccione2010}

In chemical-physics applications, having access to $\Phi(t,t_0)$ is useful well
beyond reproducing a single set of observables: it allows one to post-process
the dynamics for arbitrary initial conditions, to quantify decoherence versus
relaxation time scales, and to extract effective time-local or time-nonlocal
(memory-kernel) descriptions that connect naturally to kinetic modeling.  The
main practical bottleneck is that a direct state-by-state construction of
$\Phi(t,t_0)$ requires many independent propagations.  The following subsection
shows how the Choi representation allows one to compute the full map from a
single propagation in an enlarged space. 

For simplicity, in the following we set $t_0=0$, define $\Phi(t) = \Phi(t,0)$,
and drop the explicit dependence of the map on the initial time $t_0$.

\subsection{Choi matrix: definition and construction}
\label{subsec:choi_def}

Let $A$ be an ancilla isomorphic to $S$ with Hilbert space $\mathcal{H}_A\simeq\mathcal{H}_S$ and dimension $d_S$.  
We define the (unnormalized) maximally entangled state
\begin{equation}
|\Omega\rangle=\sum_{i=1}^{d_S} |i\rangle_S\otimes |i\rangle_A .
\end{equation}
The Choi matrix (Choi operator) associated with $\Phi$ is
\cite{Choi1975LAaiA,Jamiolkowski1972RoMP,Watrous2018}
\begin{equation}
\label{eq:choimap}
J_\Phi = \sum_{i,j} \Phi[|i\rangle\langle j|_S] \otimes |i\rangle\langle j|_A.
\end{equation}
With this convention $\operatorname{Tr}(J_\Phi)=d_S$, and the normalized ``Choi state'' is $\rho_{\mathrm{Choi}}=J_\Phi/d_S$.  
Complete positivity of $\Phi$ is equivalent to $J_\Phi\succeq 0$, and trace
preservation implies $\operatorname{Tr}_S(J_\Phi)=\mathbb{I}_A$. 

Equation~\eqref{eq:choimap} shows that $J_\Phi$ is a four-index tensor $J_\Phi(m,i;\,n,j)$ (or a $d_S^2\times d_S^2$ matrix after reshaping the paired indices $(m,i)$ and $(n,j)$).  
Once $J_\Phi$ is known, the action of the map on any system operator $\rho$ can be obtained by contraction,
\begin{equation}
\label{eq:choimapevo}
(\Phi(t)[\rho])(m,n) = \sum_{i,j} J_\Phi(m,i;n,j) \rho(i,j).
\end{equation}
This relation makes the post-processing step inexpensive: after computing
$J_{\Phi(t)}$ on a time grid, the evolution for any factorized initial condition
follows from tensor contractions in the system space.

To compute $J_{\Phi(t)}$ in practice, consider the operator
\begin{equation}
\label{eq:Y}
Y = (U\otimes I_A)(|\Omega\rangle\langle\Omega|\otimes \rho_B)(U^\dagger \otimes I_A).
\end{equation}
where $U$ is the time evolution operator of the system, defined in the Hilbert space $\mathcal{H}_S\otimes\mathcal{H}_B$.
Taking the partial trace over the environment gives
\begin{equation}
\operatorname{Tr}_B(Y) = \sum_{i,j} \operatorname{Tr}_B\!\Big[U(|i\rangle\langle j|_S\otimes \rho_B)U^\dagger\Big]\otimes |i\rangle\langle j|_A.
\end{equation}
Since by definition,
\begin{equation}
\Phi[|i\rangle\langle j|_S] = \operatorname{Tr}_B\!\Big[U(|i\rangle\langle j|_S\otimes \rho_B)U^\dagger\Big],
\end{equation}
we obtain
\begin{equation}
\label{eq:jphi}
\operatorname{Tr}_B(Y) = \sum_{i,j} \Phi[|i\rangle\langle j|_S]\otimes |i\rangle\langle j|_A \equiv J_\Phi.
\end{equation}
Thus, embedding $S$ into a maximally entangled state with an ancilla,
propagating the enlarged state under the unitary operator $U(t,t_0)$, and tracing
out the bath yields the full Choi matrix and therefore the full reduced dynamical map.  In
the next section we show how this construction can be implemented efficiently at
finite temperature by combining it with Thermofield Dynamics purification.

\section{Thermo field dynamics formulation}
\label{sec:tfd}

Equation~\eqref{eq:Y} expresses the Choi matrix starting from an operator on $S\otimes A\otimes B$.  
For a thermal environment it is convenient to rewrite this object in terms of a
\emph{pure} state by purifying the bath density matrix. To this end, introduce a tilde
(auxiliary) copy of the bath Hilbert space,
$\tilde{\mathcal{H}}_B\simeq\mathcal{H}_B$, and define the purification of
$\rho_B$ as the vector
\begin{equation}
\ket{\rho_B^{1/2}}
\equiv
\mathrm{vec}\!\left(\rho_B^{1/2}\right)
=
\sum_{k,\ell}\bigl(\rho_B^{1/2}\bigr)_{k\ell}\,|k\rangle_B\otimes|\ell\rangle_{\tilde B},
\end{equation}
so that $\rho_B=\operatorname{Tr}_{\tilde B}\bigl[\,|\rho_B^{1/2}\rangle\langle\rho_B^{1/2}|\,\bigr]$.  
With this notation, Eq.~\eqref{eq:Y} can be written as a projector
\begin{equation}
Y(t)=\ket{\phi(t)}\!\bra{\phi(t)},
\end{equation}
with
\begin{equation}
\ket{\phi(t)}
=
\bigl(U(t)\otimes I_A\otimes I_{\tilde B}\bigr)\,
\Bigl(\,|\Omega\rangle\otimes|\rho_B^{1/2}\rangle_{B\tilde B}\Bigr),
\end{equation}
and the Choi matrix follows from a partial trace over the doubled bath,
\begin{equation}
\label{eq:choitfd}
    J_\Phi = \operatorname{Tr}_{B\tilde B}[\ket{\phi(t)}\bra{\phi(t)}].
\end{equation}

TFD provides an explicit and convenient representation of
$|\rho_B^{1/2}\rangle$ for bosonic thermal
environments.\cite{TakahashiUmezawa1996IJMPB,Arimitsu1982JPSJ,Schmutz1978ZPB}  
Indeed, for an environment modeled as a
set of harmonic modes with $H_B=\sum_k \omega_k a_k^\dagger a_k$, one introduces
a thermal Bogoliubov transformation generated by
\begin{equation}
G(\beta)= i\sum_k \theta_k(\beta)\,\bigl(\tilde a_k a_k^\dagger-a_k \tilde a_k^\dagger\bigr),
\quad
\tanh\theta_k(\beta)=e^{-\beta\omega_k/2},
\end{equation}
where $\tilde a_k(\tilde a_k^\dagger)$ are annihilation (creation) operators acting on the tilde space, and $\beta$ is the inverse temperature.  
Acting with this transformation on the $T=0$ vacuum produces the purified
thermal state, $|\rho_B^{1/2}\rangle=e^{iG(\beta)}|0\rangle_{B\tilde B}$.

For numerical propagation it is convenient to shift the temperature dependence
from the initial state to the
Hamiltonian.\cite{BorrelliGelin2021WCMS,GelinBorrelli2017AdP,BorrelliGelin2016JCP}
Because the partial trace in
Eq.~\eqref{eq:choitfd} is invariant under unitary transformations acting only on
$B\tilde B$, 
\begin{equation}
\label{eq:choitfd2}
    J_\Phi = \operatorname{Tr}_{B\tilde B}[\ket{\phi(t)}\bra{\phi(t)}]
     = \operatorname{Tr}_{B\tilde B}[\ket{\varphi(t)}\bra{\varphi(t)}].
\end{equation}
where, for a time-independent Hamiltonian $H$, 
\begin{equation}
\label{eq:evotfd}
    \ket{\varphi(t)} = e^{-i H_\theta t}\ket{\Omega}\ket{0}_{B\tilde B}
\end{equation}
where $H_\theta=e^{-iG}He^{iG}$
is a transformed Hamiltonian operator.  
Therefore, the basic ingredient to compute the 
Choi map for a quantum system at temperature $T$, is the evolution of 
the state  $\ket{\Omega}\ket{0}_B$ 
with the temperature-dependent Hamiltonian $H_\theta$.  
For linear vibronic-coupling models, $H_\theta$ can be expressed in a compact
form\cite{TakahashiEtAl2024TJoCP} 
\begin{equation}
\label{eq:htheta}
\begin{aligned} 
H_{\theta}= & \sum_{n, m}\left(\varepsilon_{n m} +\sum_{k=1}^{N_B} \frac{g_{k n m}(\beta)}{\sqrt{2}}\left(b_k^{\dagger}+b_k\right)\right)|n\rangle\langle m| \\ 
& +\sum_{k=1}^{N_B} \omega_k b_k^{\dagger} b_k.
\end{aligned}
\end{equation}
Here, $N_B$ denotes the total number of physical and tilde 
vibrational DoFs,  with corresponding annihilation and creation operators 
 $b_k$ and $b_k^\dagger$. The effective couplings $g_{knm}(\beta)$ include the thermal factors
arising from the Bogoliubov transformation.  
The frequencies $\omega_k$ are allowed to take positive and
negative values
so that physical and tilde modes can be treated within a unified notation.\cite{TakahashiEtAl2024TJoCP,BorrelliGelin2017SR,TamascelliEtAl2019PRL}

Equations~
\eqref{eq:choitfd}--\eqref{eq:evotfd}--\eqref{eq:htheta} provide the working recipe for
the following sections: a single TT propagation of
$|\Omega\rangle\otimes|0\rangle$ under $H_\theta$ yields the full reduced
dynamical map (via its Choi matrix) on a chosen time grid.  

\section{Static disorder and ensemble-averaged maps}
\label{sec:disorder}

Experimental observables for large molecular systems, in particular aggregates 
and materials, often reflect an
\emph{ensemble average} over slightly different realizations of the system
Hamiltonian (site-energy disorder, coupling disorder).  
This is commonly referred to as static disorder.
Let $\sigma$ denote a set of static parameters characterizing a realization, with probability density $p(\sigma)$.  
For each realization one obtains a reduced map $\Phi(t;\sigma)$ and a reduced density $\rho_S(t;\sigma)=\Phi(t;\sigma)[\rho_S(0)]$.  
Because $\Phi$ acts linearly on the system density matrix, the ensemble-averaged reduced state can be written as
\begin{equation}
\bar\rho_S(t)=\int d\sigma\,p(\sigma)\,\rho_S(t;\sigma)
=
\bar\Phi(t)\big[\rho_S(0)\big],
\end{equation}
where
\begin{equation}
\bar\Phi(t)=\int d\sigma\,p(\sigma)\,\Phi(t;\sigma).
\end{equation}
Thus, static disorder averaging can be performed directly at the map level.

Following Ref.~\onlinecite{GelinEtAl2021JCP}, static disorder can be embedded into the
Hamiltonian by introducing auxiliary bosonic modes with \textit{zero frequency} in the Hamiltonian Eq.~\ref{eq:htheta} leading to the operator
\begin{equation}
\bar H_\theta
=
H_\theta
+
\sum_{n,m}
V_{nm}\,\frac{\sigma_{nm}}{\sqrt{2}}\left(z_{nm}^{\dagger}+z_{nm}\right),
\label{eq:htheta_disorder}
\end{equation}
where $V_{nm}=|n\rangle\langle m|$ is an electronic operator, $\sigma_{nm}$ sets
the disorder strength of the corresponding Hamiltonian parameter, $z_{nm}$
are auxiliary bosonic annihilation operators, and $D$ is the number of system parameters subject to static disorder.  With an
appropriate choice of the initial state of the $z_{nm}$ modes (encoding the
target distribution $p(\sigma)$), propagation under $\bar H_\theta$ yields the
disorder-averaged map $\bar\Phi(t)$ from a \emph{single} simulation, without
explicit sampling over many disorder realizations.\cite{GelinEtAl2019JPCL}
We note that the additional modes do not constitute a thermal bath in
the usual meaning: because their frequency is zero they cannot accept energy
and therefore they do not contribute to any energy flow into
the environment.

\section{Tensor Train Representation}
\label{sec:tt_rep}

High-dimensional TFD states can be described in a controlled approximate manner using numerical tensor network techniques. In our work, we employ the TT format.\cite{Oseledets2011SJSC,White1992PRL}  
Besides providing a compact representation, the TT format is convenient for the evaluation of partial traces and is therefore well suited to the construction of reduced dynamical maps at finite temperature.
Below, we summarize the TT representation used in this work.

Define $L = D + N_B + 2$
as the total number of degrees of freedom of the TFD state. 
The generic TFD state can then be written in TT form as
\begin{equation}
\label{eq:phitt}
|\varphi\rangle
=
\sum_{i_1,\ldots,i_L}
G^{(1)}(i_1)\cdots G^{(L)}(i_L)\,
|i_1\rangle \cdots |i_L\rangle .
\end{equation}
Here, $\{ |i_k\rangle \}_{i_k=1}^{p_k}$ is the truncated basis associated with the $k$th degree of freedom. The $k$th TT core is a three-index tensor
\begin{equation}
G^{(k)} \in \mathbb{C}^{r_{k-1}\times p_k\times r_k},
\quad
G^{(k)}_{\alpha_{k-1},\,i_k,\,\alpha_k}
=
\bigl[G^{(k)}(i_k)\bigr]_{\alpha_{k-1},\,\alpha_k},
\end{equation}
where $r_k$ are the TT ranks 
(with boundary values $r_0=r_L=1$). In a TT representation, the ordering of the degrees of freedom strongly affects both the accuracy and the computational efficiency of the approximation. In the present work, we use the ordering
\begin{equation}
z_1 - \cdots - z_D - A - S - b_1 - \cdots - b_{N_B},
\label{eq:tt_seq}
\end{equation}
where $A$ and $S$ denote the ancilla and system degrees of freedom, $z_i$ the auxiliary modes associated with static disorder, and $b_j$ the bath vibrational modes,
ordered according to increasing absolute values of the frequency $|\omega_{j}|$.
Finally, the partial trace over the environmental degrees of freedom in Eq. \ref{eq:choitfd} can be written as
\begin{equation}
J_{\Phi}(i_A,i_S;j_A,j_S)
=
\sum_{\mathbf{i}_{\mathrm{env}}}
\varphi_{\mathbf{i}_{\mathrm{env}},\,i_A,\,i_S}
\varphi^{*}_{\mathbf{i}_{\mathrm{env}},\,j_A,\,j_S},
\end{equation}
where  $\mathbf{i}_{\mathrm{env}}$ collects all environmental indices.

For the sake of completeness, we report the explicit 
formulae for an efficient evaluation of the trace 
in the appendix.

%===========================================================
\section{Numerical application: Exciton transport in the FMO complex}
\label{sec:numerics}

We illustrate the proposed TFD--Choi-TT framework on a model
Fenna--Matthews--Olson (FMO) complex, focusing on exciton transport within
subunit A of the FMO trimer having eight bacteriochlorophylls (BChls) \cite{SchmidtamBuschEtAl2011JPCL,MoixEtAl2011JPCL}.  
In the last decade this system has become a reference both for understanding coherence effects in quantum biology\cite{ScholesEtAl2017N} and for the development of new quantum dynamical methodologies.\cite{PriorEtAl2010PRL,SchulzeEtAl2016JCP,BorrelliGelin2017SR}
We employ a realistic linear vibronic-coupling description in which the excitonic Hamiltonian is adopted from Moix \textit{et al.}\cite{MoixEtAl2011JPCL} and each site is coupled to
a structured spectral density derived by Coker and co-workers.\cite{LeeCoker2016JPCL,RiveraEtAl2013JPCB}  
The system Hamiltonian is reported in the the SM file.
The spectral densities are discretized using the low-rank
interpolative-decomposition (ID) strategy developed by the authors.\cite{TakahashiBorrelli2024JCP,TakahashiEtAl2024TJoCP} 
The ID approach enables us to 
build a minimal discrete bath model that still reproduces the target open-system dynamics within a chosen time window 
$[0,T_c]$ at a given temperature $\beta$.
The parameters $g_{knm}(\beta)$ and the frequencies $\omega_k$ are obtained 
by a low-rank discretization of the bath correlation function, as explained in Refs. 
\onlinecite{TakahashiBorrelli2024JCP,TakahashiEtAl2024TJoCP}

In the discretization process we set a 
temperature of 300 K and an 
upper time limit $T_c = 800$ fs.
This limit sets the range of validity of our quantum dynamical calculation because after that the bath correlation function looses the requested 
accuracy.
The eight discretized forms of the SDs are given in the SI.

\begin{figure}[t]
  \centering
  \includegraphics[width=\linewidth]{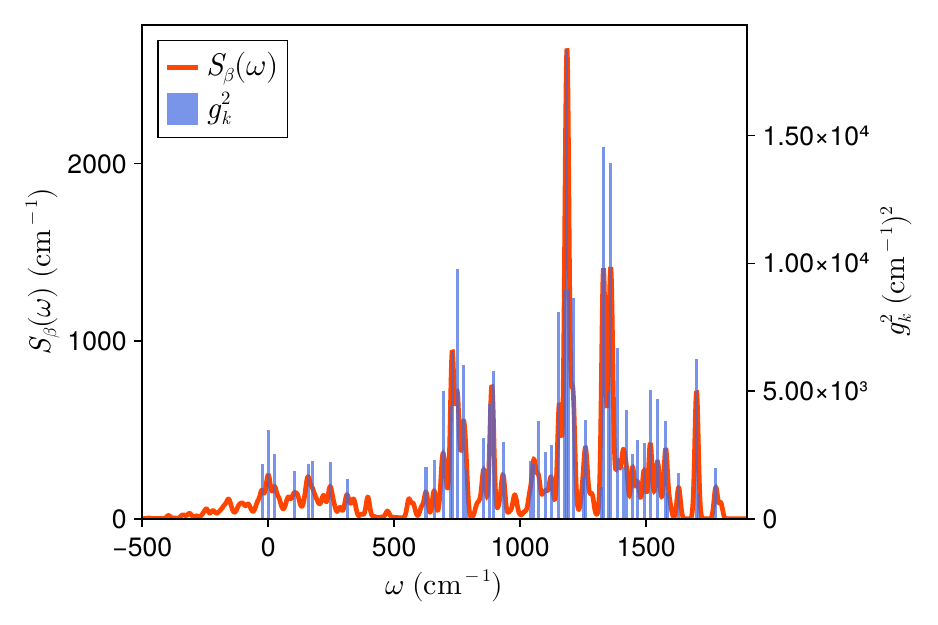}
\caption{Quantum noise spectral density and its interpolative-decomposition
discretizations for site 1 of the subunit A of the FMO complex. The solid curve shows the
continuous function from MD, while the sticks indicate 
the discrete frequencies and couplings employed in the 
TT propagation (50 effective modes).}
\label{fig:SDBChl12}
\end{figure}

Figure~\ref{fig:SDBChl12} shows a representative discretization of the quantum noise spectral density (QNSD), 
  $S_\beta(\omega) = J(\omega)\left[1+\coth(\beta\omega/2)\right]$,
for BChl 1. 
Here, $J(\omega)$ is the spectral density. See SI for more details.
The red continuous line is obtained directly from MD simulations as explained in Ref. \onlinecite{LeeCoker2016JPCL}. The blue sticks represent the discretized form. 
For the present model, the overall TFD Hamiltonian at $T=300$~K
comprises 411 vibrational degrees of freedom (physical and tilde modes
combined).

The time evolution in the TT representation is computed using the time-dependent 
variational principle (TDVP) technique for TTs, in its KSL formulation.\cite{LubichEtAl2015SJNA}
This integration technique keeps the ranks of the TT cores frozen 
during the evolution, therefore, several runs with increasing ranks are required to check the convergence of the
calculation. 
We note that the linear vibronic coupling Hamiltonian of Eq. 
\ref{eq:htheta} has a star-like 
system-bath connectivity and TDVP becomes mandatory to treat long-range interactions in the TT format. Hence we chose to adopt the one-site TDVP methodology.
Furthermore, while other methodologies can be applied\cite{Dolgov2019CMAM,BorrelliDolgov2021JPCB,TakahashiBorrelli2025JCP}
TDVP has a relatively low computational cost, and is norm-preserving by construction.

\subsection{Reduced map construction and density propagation}

From a single propagation of the thermofield initial state
  $|\Omega\rangle\otimes|0\rangle$, with the Hamiltonian operator $H_\theta$ [Eq.~\eqref{eq:htheta}], we
compute the Choi matrix $J_{\Phi(t)}$ on a discrete time grid by the TT
contraction scheme described in the Appendix~\ref{sec:tt_trace}.  

In the model under investigation $d_S=8$, $J_{\Phi(t)}$ is
a $d_S^2\times d_S^2=64\times 64$ matrix, and the normalized Choi state is
$\rho_{\mathrm{Choi}}(t)=J_{\Phi(t)}/d_S$.
Once $J_{\Phi(t)}$ is stored, the reduced dynamics for arbitrary factorized
initial conditions follows from the contraction in Eq.~\eqref{eq:choimapevo} 
at negligible additional computational cost.

First, inspection of Eq. \ref{eq:choitfd} shows immediately that the reduced map is completely positive (CP) by construction. Trace preservation (TP), however, requires a distinct condition. In the Choi representation, TP is equivalent to
\begin{equation}
\Phi \text{ is TP } \Longleftrightarrow \Tr_S J_\Phi = \mathbb{I}_A.
\end{equation}
By contrast, the TDVP propagation in TT form guarantees only preservation of the norm of the full purified state $\langle \varphi(t) | \varphi(t) \rangle = 1$.
This does not, in general, imply exact trace preservation of the reduced map. Indeed, norm conservation is a scalar constraint on the global TT state, whereas TP is a matrix condition on the ancilla marginal of the Choi state. As a consequence, once the dynamics is projected onto a finite-rank TT manifold, exact TP of the reduced map is not guaranteed \textit{a priori}.

Nevertheless, we have verified that the average deviation from TP 
decreases very rapidly by decreasing the time-step going from $1.3\times 10^{-4}$ at $\Delta t = 1$ fs to $2.7\times 10^{-7}$ at $\Delta t = 0.5$ fs and
$4.0\times 10^{-8}$ at $\Delta t=0.25$ fs. This suggests that the TP residual is mostly limited by finite-time-step discretization errors.
Further, we have found that it is almost rank independent, 
and remains constant  throughout the dynamics  (see SI). 
These results support both the numerical accuracy of the results and the correctness of the actual implementation.

\begin{figure}[t]
  \centering
  \safeincludegraphics[width=0.9\linewidth]{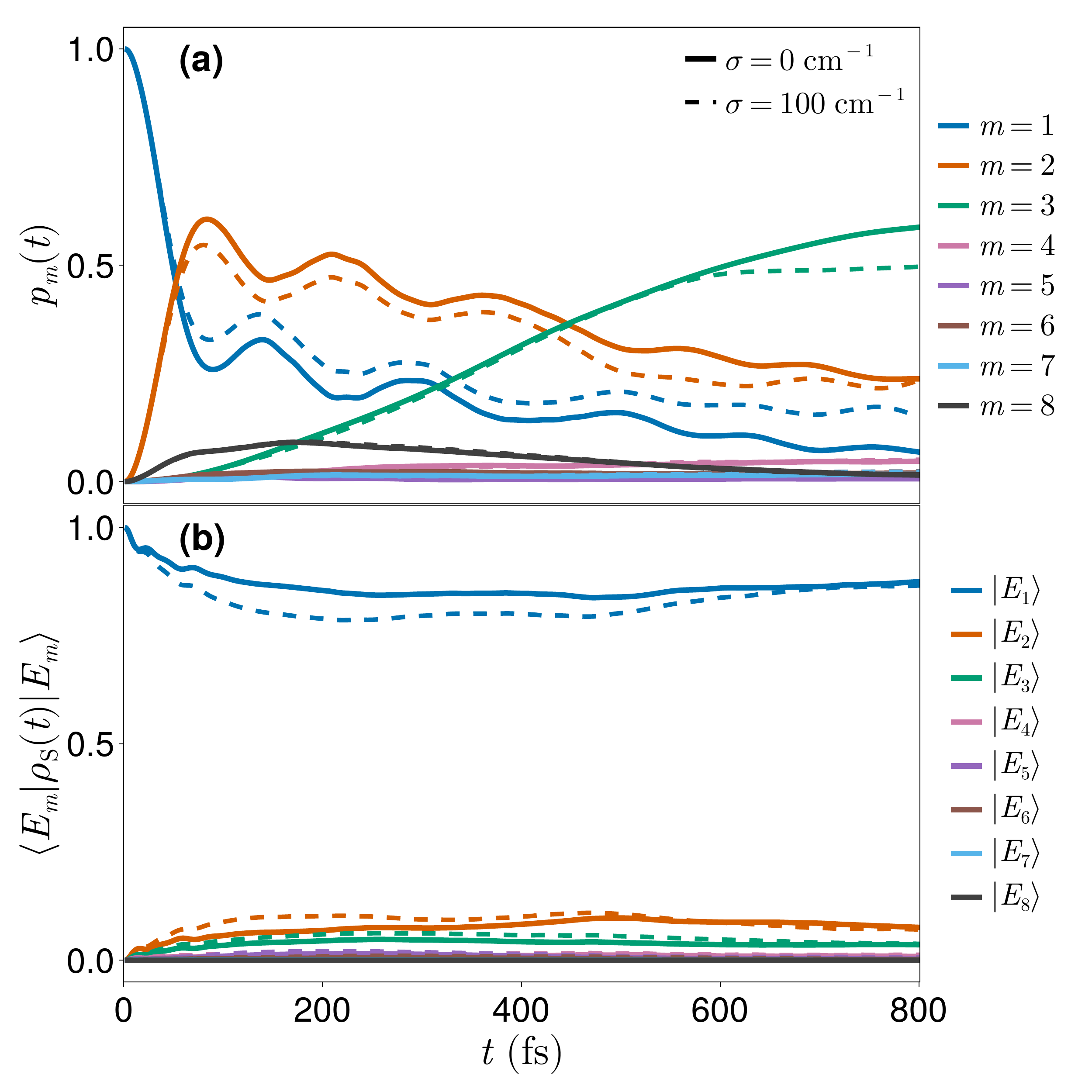}
  \caption{Panel a: full line, site populations $p_m(t)=\langle m|\rho_S(t)|m\rangle$ for the
  eight-site FMO model starting from an initial excitation localized on site~1; dashed line, same dynamics but with a diagonal static disorder $\sigma=100$ cm$^{-1}$.
  Panel b: Excitonic state populations, $\mel{E_m}{\rho_S(t)}{E_m}$, for the same model, for $\ket{E_1}$ as initial state;
  dashed line, same dynamics but with a diagonal static disorder $\sigma=100$ cm$^{-1}$.}
  \label{fig:mapevo}
\end{figure}

An immediate  application of the map is the time propagation of a specific state of the system.
Fig.~\ref{fig:mapevo}a) presents the site populations
$p_m(t)=\mel{m}{\rho_S(t)}{m}$ as a function of time as obtained from Eq. \ref{eq:mapevo}, for an initial condition in which the excitation is fully localized on site 1. 

To the best of our knowledge, this is the first fully quantum calculation at nonzero temperature for the eight-site FMO model employing site-specific spectral densities. 
While our results are broadly consistent with previous studies\cite{RiveraEtAl2013JPCB}, a direct quantitative comparison is not possible because different density-matrix propagation methods were used.

The effect of static disorder can be introduced by using the Hamiltonian in Eq.~\ref{eq:htheta_disorder} with properly chosen dispersion parameters $\sigma_{nm}$.
Dashed lines in Fig.~\ref{fig:mapevo}a) represent the population dynamics 
with a diagonal (site energy only) static disorder, $V_{nn} = \ket{n}\bra{n}$, and $\sigma = 100$ cm$^{-1}$ for all sites, when the excitation is initially localized on site 1.
We clearly see that the populations of sites 1, 2 and 3 are strongly affected by the disorder. After 600 fs the occupation of site 3 appears to become almost stationary.
We note that in this calculation the number of sites in the chain increases and convergence of the calculation with respect to the ranks of the TT cores becomes more difficult.
To keep the computational cost at a reasonable level we used a maximum TT rank of 220. 

We next consider excitation by a laser pulse that prepares a coherent superposition of electronic states. For definiteness, we assume that the pulse populates the first excitonic state, $\ket{E_1}$, namely the lowest-energy eigenstate of the electronic Hamiltonian. By expressing the maps in the excitonic basis, we can straightforwardly evaluate the excitonic populations, $\mel{E_i}{\rho_S(t)}{E_i}$,
as functions of time; the corresponding dynamics are shown in Fig.~\ref{fig:mapevo}b).

Our aim here is to show the capabilities of the methodology, nonetheless we note 
that in the excitonic basis the population transfer is much less pronounced. Dashed lines show the 
effect of static disorder on electronic exciton populations.
At variance with the previous case we see that static disorder seems to have only minor effects on the excitonic dynamics.

\subsection{Choi-spectrum diagnostics: decoherence versus relaxation}

\begin{figure}[t]
  \centering
  \safeincludegraphics[width=0.9\linewidth]{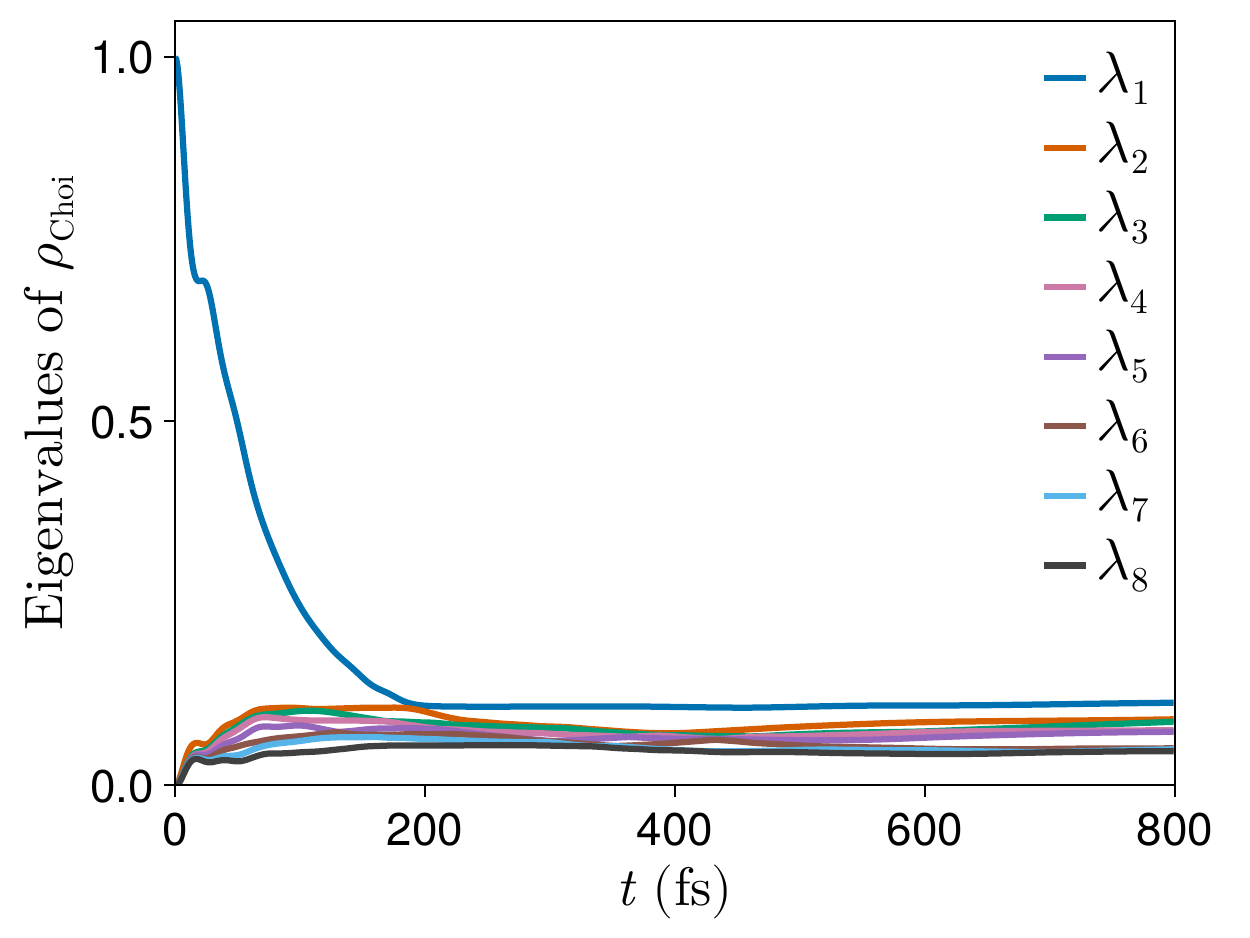}
  \caption{Leading eigenvalues $\{\lambda_\alpha(t)\}$ of the normalized Choi
  state for the
  eight-site FMO model.} \label{fig:eigs}
\end{figure}

Beyond straightforward time propagation of the system density matrix, the Choi representation allows one to
characterize the dynamical map itself.  Figure~\ref{fig:eigs} displays the
leading eigenvalues $\{\lambda_\alpha(t)\}$ of the normalized Choi state
$\rho_{\mathrm{Choi}}(t)$,
providing information on the degree of system-bath entanglement.
At short times the spectrum is dominated by a single eigenvalue close to one,
indicating that the reduced dynamics is nearly unitary.  
As time increases, weight is transferred from this dominant component to several
subleading eigenvalues, reflecting the growth of system--bath correlations and
the increasing complexity of the reduced evolution.

The spectral evolution suggests two distinct stages.  An initial fast stage
features a rapid drop of the leading eigenvalue and a rise of a few additional
eigenvalues, consistent with a decoherence time scale over which coherences are
strongly suppressed.  A second, slower stage gradually populates a broader tail
of eigenvalues, reflecting population relaxation and redistribution among
excitonic states.  In this way, the Choi spectrum provides a compact,
basis-independent view of how noise builds up over time in the reduced dynamics.

\subsection{Entropy and effective rank}

A complementary global measure of the Choi-state mixing is the von Neumann entropy
\begin{equation}
S(t)=-\operatorname{Tr}\!\left[\rho_{\mathrm{Choi}}(t)\log_2\rho_{\mathrm{Choi}}(t)\right]
= -\sum_\alpha \lambda_\alpha(t)\log_2\lambda_\alpha(t).
\end{equation}

The quantity $S(t)$ therefore measures how broadly the eigenvalue weight of the Choi state is distributed.
A value $S(t)=0$ corresponds to a pure
Choi state, i.e. a single-Kraus, unitary reduced evolution for a
trace-preserving map on equal input and output spaces.  Increasing entropy indicates 
that more Kraus components are required to represent the reduced
dynamics, reflecting the loss of information from the system into the bath.

For a $d_S$-level system, the Choi state acts on a $d_S^2$-dimensional
operator space, and hence
\begin{equation}
0 \leq S(t) \leq S_{\max}=\log_2(d_S^2)=2\log_2 d_S.
\end{equation}
The upper bound would correspond to a maximally mixed Choi state, for which
all $d_S^2$ Choi eigenvalues are equally populated.  In the present FMO
calculation the entropy remains below this value, showing that the bath
does not fully randomize the reduced dynamics over the accessible
operator space.

It is also useful to summarize the spectrum by an effective rank,
\begin{equation}
r_{\mathrm{eff}}(t)=2^{S(t)},
\end{equation}
which estimates the number of eigenvalues that are effectively populated:  
if exactly $k$ eigenvalues were equally weighted, then
$r_{\mathrm{eff}}=k$.
Thus $r_{\mathrm{eff}}(t)$ gives an intuitive estimate of the number of dynamically relevant Kraus directions.

\begin{figure}[t]
  \centering
  \safeincludegraphics[width=0.9\linewidth]{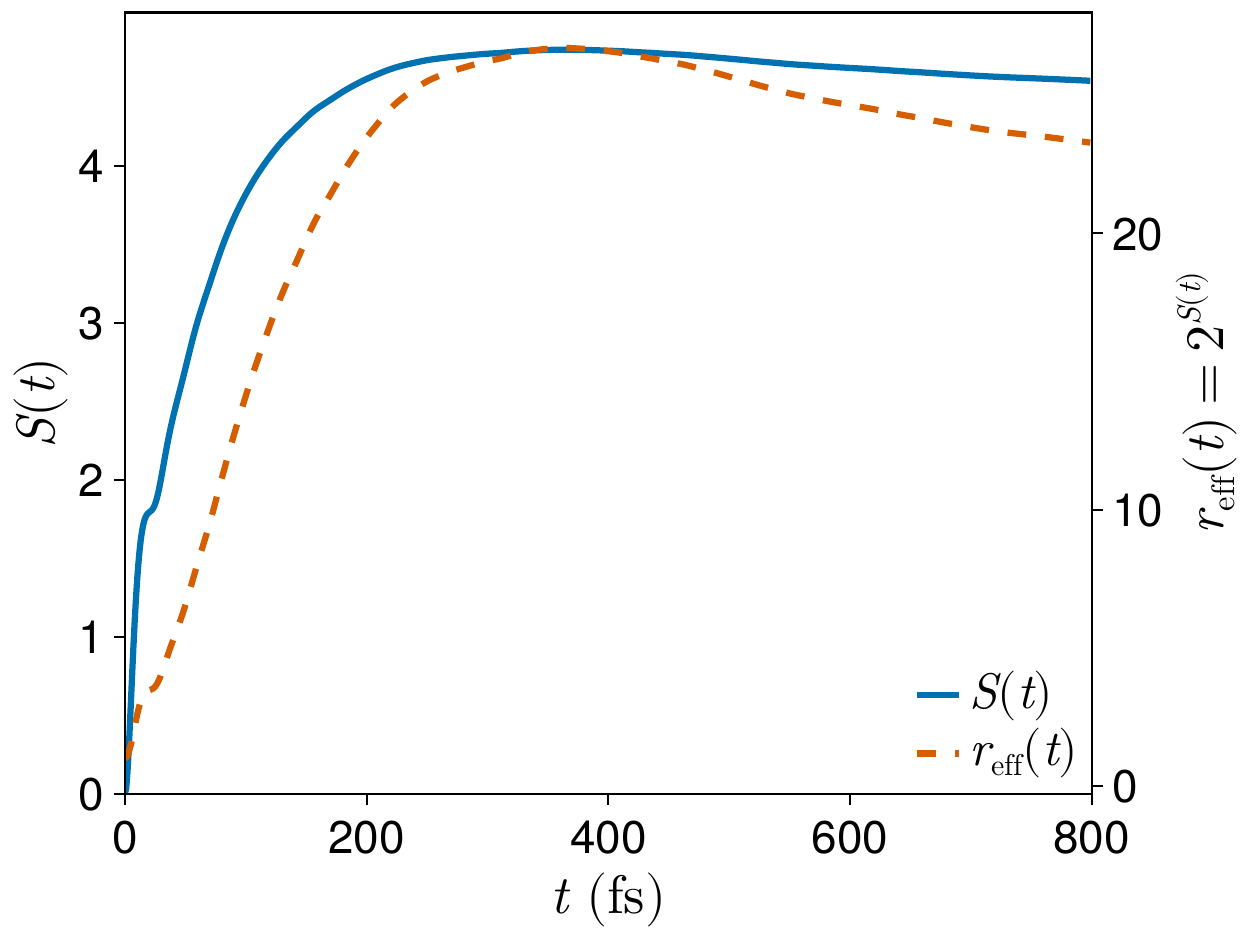}
  \caption{Von Neumann entropy $S(t)$ of the normalized Choi state
  $\rho_{\mathrm{Choi}}(t)$ (left axis) and effective rank
  $r_{\mathrm{eff}}(t)=2^{S(t)}$ (right axis). 
  } \label{fig:entropy}
\end{figure}

Figure~\ref{fig:entropy} shows $S(t)$ and $r_{\mathrm{eff}}(t)$ for the
current FMO model.  Both quantities rise sharply at early times, indicating
the rapid onset of system--bath entanglement and the corresponding departure
from nearly unitary reduced dynamics.  This initial increase is consistent
with the fast redistribution of weight observed in the Choi eigenvalue
spectrum.  At intermediate times, the entropy continues to grow more slowly,
reflecting additional population relaxation and bath-induced mixing of the
reduced map.  The maximum occurs on the few-hundred-femtosecond timescale,
where the effective rank reaches its largest value and, in quantum information language, the channel samples
the broadest set of operator-space modes.

At later times, both $S(t)$ and $r_{\mathrm{eff}}(t)$ decrease slightly rather
than approaching the maximally mixed limit.  This behavior indicates that the
long-time reduced dynamics is not simply becoming more random.  Instead, the
map relaxes toward a stationary regime in which a smaller subset
of Choi eigenmodes carries most of the spectral weight.  The entropy and
effective-rank analysis therefore support the eigenvalue-level picture:
the environment rapidly generates a high-rank, non-unitary reduced dynamics,
but the asymptotic channel retains significant structure and does not become
fully depolarizing.

\subsection{Memory effects, time-local generators, and long-time extrapolation}
\label{subsec:markovianity}

A most interesting aspect of having access to the full map sequence $\{\Phi(t_n)\}$ 
is the possibility to directly
assess memory effects and to construct reduced descriptions that
are particularly useful for long-time dynamics and kinetics.
Two complementary objects are
the time-local generator $\mathcal{L}(t)$\cite{TokuyamaMori1976PTP,TokuyamaMori1976PTPa,BreuerEtAl2001AoP} and the
Nakajima--Zwanzig (NZ) memory kernel.\cite{Zwanzig1954JCP,Nakajima1958PoTP,Mori1965PTP}
Both have been widely used in the past to describe the reduced density matrix 
dynamics, but their determination can be very challenging.

\paragraph*{Time-local (TL) generator.}
As already described in Section \ref{sec:choi},
the TL generator (also referred to as time-convolutionless, TCL)
obtained from $\Phi(t)$, i.e.\ the superoperator
\begin{equation}
\mathcal{L}(t)=\dot{\Phi}(t)\,\Phi(t)^{-1},
\label{eq:TCL_generator}
\end{equation}
defines the time-local master equation
$\dot{\rho}_S(t)=\mathcal{L}(t)\rho_S(t)$.

In principle, $\mathcal{L}(t)$ can be numerically evaluated from discrete map data by
representing $\Phi(t)$ as a $d_S^2\times d_S^2$ matrix in Liouville space
and using finite differences, 
\begin{equation}
    \mathcal{L}(t_{n+1})\approx
    \frac{\Phi(t_{n+1})-\Phi(t_n)}{\Delta t}\circ \Phi(t_n)^{-1}.
\end{equation}
where $\Delta t=t_{n+1}-t_n$ and the inverse is taken on the numerically
relevant subspace (regularized or as a pseudoinverse if needed).\cite{NestmannEtAl2021PRX} 
A crossover to
an effectively time-local Markovian description is indicated by a slow variation of
  $\mathcal{L}(t)$ in time, corresponding to the
establishment of approximately constant relaxation rates after the bath
correlation time.\cite{Contreras-PulidoEtAl2012PRB,KarlewskiMarthaler2014PRB}

In practice, however, the superoperator obtained from the
Choi matrix can become ill-conditioned at intermediate and long times:
dissipation leads to small singular values of $\Phi(t)$, and the
combination of finite-difference derivatives with a numerical inverse can
amplify noise and produce spurious spikes in individual generator
elements, making Eq. \eqref{eq:TCL_generator} hardly
usable.\cite{BhattacharjeeEtAl2023} These apparent divergences are a known
feature of TL reconstructions of the
reduced dynamics.\cite{SayerMontoya-Castillo2023JCP} 
Possible strategies that can help overcome these difficulties include either regularizing the inverse (e.g.\ using truncated-SVD or
 pseudoinverse) or using a discrete time form of Eq. \ref{eq:TCL_generator}. 
\cite{SayerMontoya-Castillo2023JCP,BhattacharyyaEtAl2025TJoCP}
Considering the above numerical issues, we analyze an alternative approach to
reduced dynamics via the time non-local NZ approach, which, as we shall see
below, is more robust and stable.

\paragraph*{NZ memory kernel and transfer tensors.}

Using the relation $\rho_S(t)=\Phi(t)\rho_S(0)$, it can be verified that
the NZ equation in the interaction picture for the reduced density matrix
\begin{equation}
\dot{\rho}_S(t)=\int_0^t \mathcal{K}(t-s)\,\rho_S(s)\,ds
\label{eq:NZ_rho}
\end{equation}
is equivalent to the Volterra equation for the map,
\begin{equation}
\dot{\Phi}(t)=\int_0^t \mathcal{K}(t-s)\,\Phi(s)\,ds,\qquad \Phi(0)=\mathcal{I}.
\label{eq:NZ_map}
\end{equation}
Here the NZ kernel $\mathcal{K}(t)$ contains information about past evolution 
of the system. 
Discretizing Eq.~\eqref{eq:NZ_map} on a uniform grid $t_n=n\Delta t$ using a
forward difference for $\dot\Phi(t_n)$ and a left-rectangle rule for the
convolution gives, for $n\ge 1$, \begin{equation}
\frac{\Phi_{n+1}-\Phi_n}{\Delta t}
=
\Delta t\sum_{m=0}^{n-1}K_{n-m}\,\Phi_m,
\qquad n\ge 1,
\label{eq:disc_map}
\end{equation}
and the recursion
\begin{equation}
K_n
=
\frac{\Phi_{n+1}-\Phi_n}{\Delta t^2}
-
\sum_{j=1}^{n-1}K_j\,\Phi_{n-j},
\qquad n\ge 1.
\label{eq:kernel_rec}
\end{equation}
A detailed and enlightening analysis of the approximations involved in the
above discretization, as well as a discussion of more sophisticated numerical approaches
have recently been provided by Makri.\cite{Makri2025JCTC} 
See also Sayer and Montoya-Castillo\cite{SayerMontoya-Castillo2024JCP} for a related
approach to the numerical determination of the NZ kernel.
We mention that, while other methods for the direct computation of the NZ kernel
based on projection techniques are available, their implementation can be rather
complex, especially if an exact representation is sought.\cite{ShiGeva2003JCP,ZhangEtAl2006JCP,Montoya-CastilloReichman2016JCP,Montoya-CastilloReichman2017JCP,KellyEtAl2016JCP,PfalzgraffEtAl2019JCP}

Once the kernel is known, reduced states can be propagated by the discrete NZ equation,
\begin{equation}
\rho_{n+1}=\rho_n+\Delta t^2\sum_{m=0}^{n-1}K_{n-m}\,\rho_m.
\label{eq:disc_rho_NZ}
\end{equation}

The same information can be cast into the transfer-tensor form of Cerrillo and
Cao,\cite{CerrilloCao2014PRL,BuserEtAl2017PRA,SayerMontoya-Castillo2024JCP,PollockModi2018Q} 
\begin{equation}
\Phi_n=\sum_{m=1}^{n}T_m\,\Phi_{n-m},\qquad n\ge 1,\qquad \Phi_0=\mathbb I.
\label{eq:ttm_id}
\end{equation}
which yields the propagation rule
\begin{equation}
\rho_n=\sum_{m=1}^{n}T_m\,\rho_{n-m}.  
\end{equation}

{Most importantly, if the bath correlation time is finite, the kernel (or transfer tensors) 
falls off as a function of the time lag,
the sum can be truncated at a finite memory length $K <  n$, 
and the long-time dynamics can be in principle
extrapolate beyond the original simulation window.
This is equivalent to stating that $\Phi(t)$ is a $K$th order autoregressive process,\cite{PercivalWalden1993} whereby the process at time $t_n$ is described as a linear combination of the $K$ processes at times $t_{n-1}\ldots t_{n-K}$.}

\begin{figure}[t]
  \centering
  \safeincludegraphics[width=0.9\linewidth]{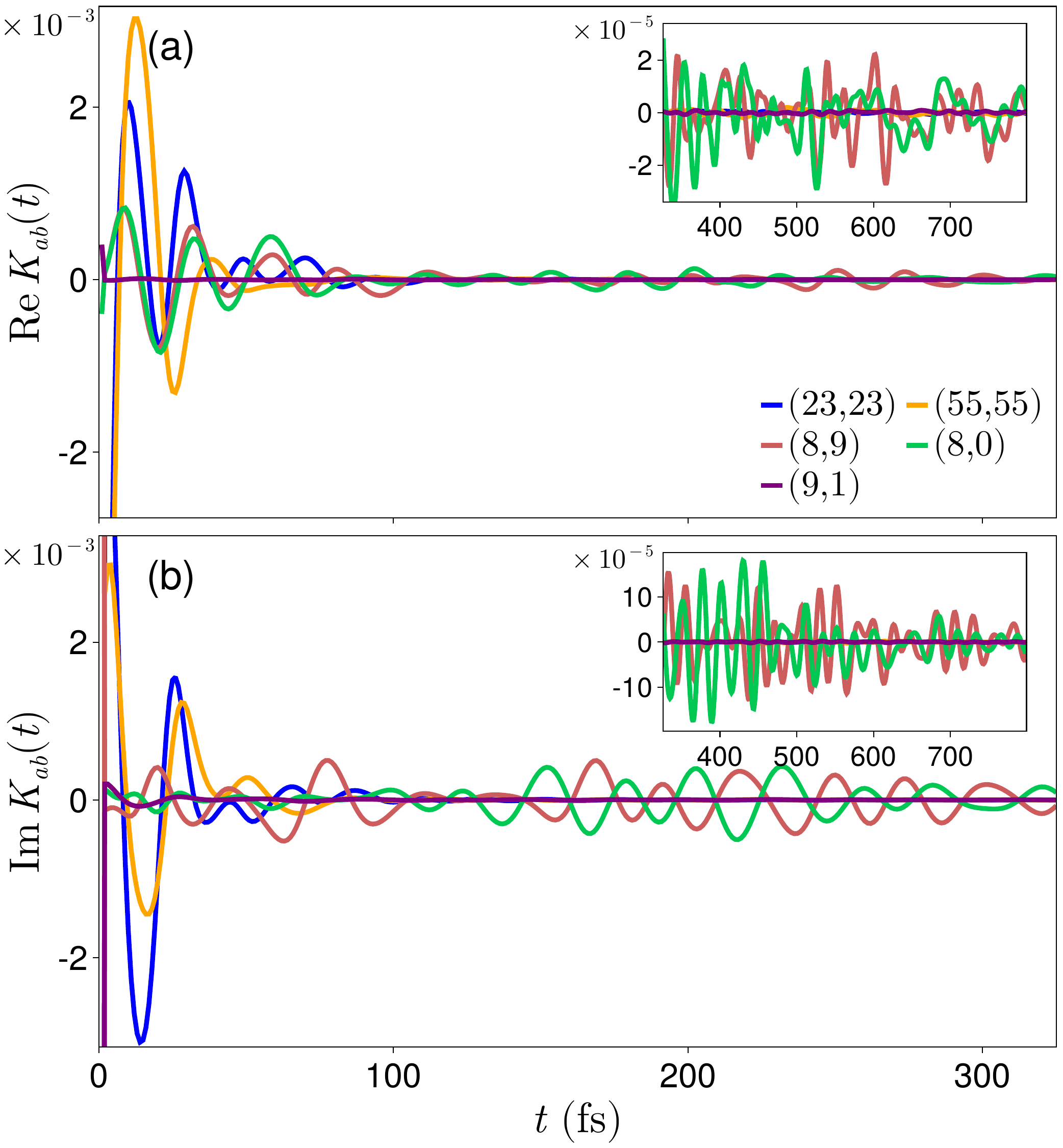}
  \caption{Representative components of the Nakajima--Zwanzig memory kernel
  $\mathcal{K}(t)$ reconstructed from the discrete map sequence $\{\Phi_n\}$ via
  Eq.~\eqref{eq:kernel_rec}.  The y-axis is scaled by $10^{-3}$ in the main plot and by $10^{-5}$ in the inset.  Selected kernel elements are shown.}
  \label{fig:computed_kernel}
\end{figure}

\begin{figure}[t]
  \centering
  \safeincludegraphics[width=0.9\linewidth]{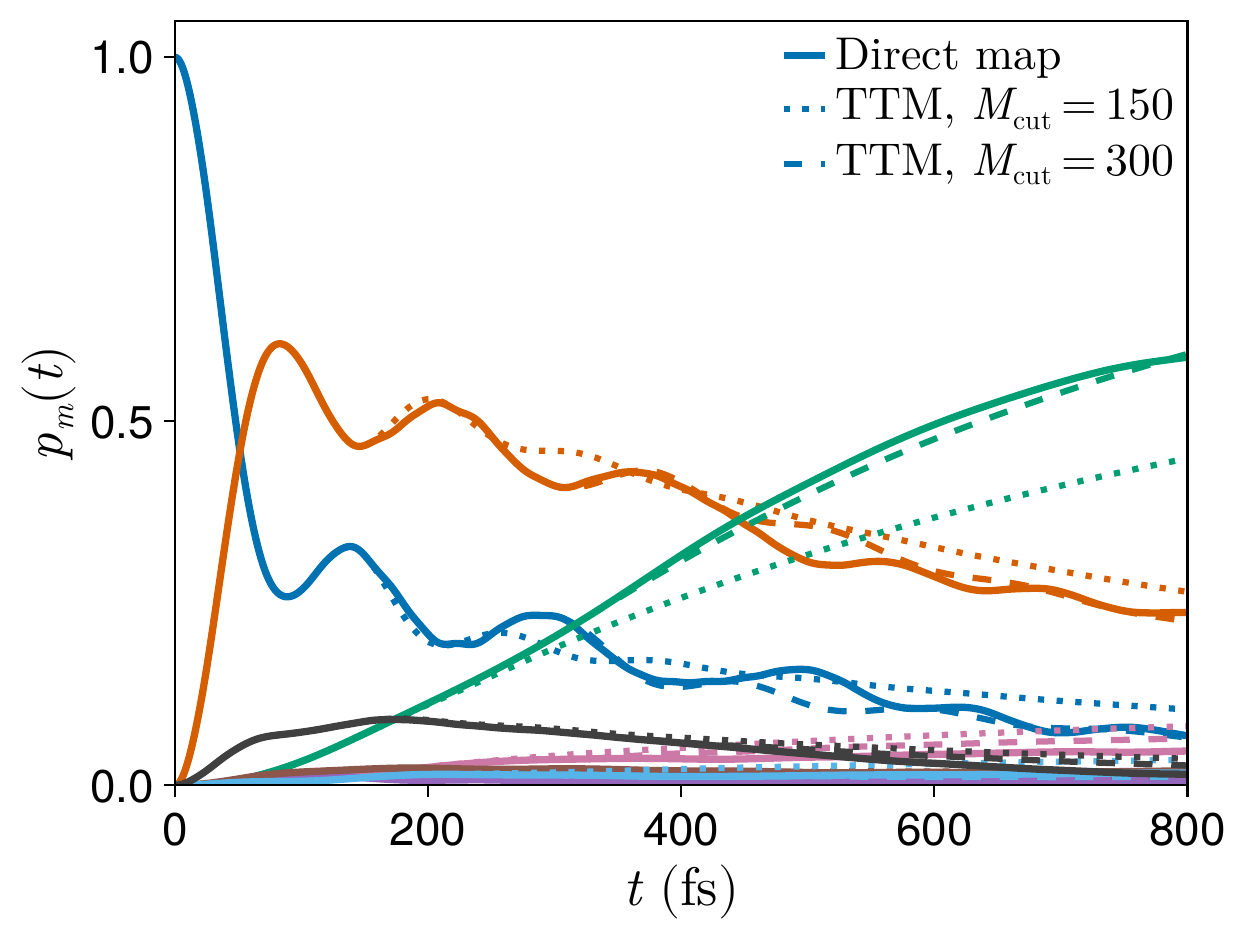}
  \caption{
  Site populations for the
  eight-site FMO model starting from an initial excitation localized on site~1.
  Solid lines: direct propagation with the reduced map $\Phi(t)$. 
  Dotted lines: TTM approximation with memory length  $M=150$~fs; 
  Dashed lines: TTM approximation with memory length  $M=300$~fs.
  }
  \label{fig:ttm_approx_population}
\end{figure}

Figure~\ref{fig:computed_kernel} shows representative components of the
reconstructed memory kernel.  
We note that the four-index notation for the kernel elements has been flattened into a two-index matrix notation; the two notations are, of course, interchangeable. The kernel elements decay rapidly on the scale of a few
hundred femtoseconds, providing a quantitative estimate of the memory time in
this model and motivating finite-memory extrapolation.  However, we note that 
oscillations still persist at long times indicating a incomplete dephasing and relaxation
of the bath. This makes non-Markovian effects non-negligible at least up to 800 fs.
Figure \ref{fig:ttm_approx_population}
demonstrates the resulting transfer-tensor reconstruction and finite-memory
extrapolation of site populations. In both cases the initial density matrix $\rho_S(0)$ 
describes a state fully localized on site 1. Clearly, any other initial state would work. 
As expected, when the memory cutoff is too short, the NZ/TTM approach is only approximate, although the main features are already recovered with a 300 fs memory window.
This is possible because the memory kernel is time-translationally invariant, i.e. it depends only on the time lag. Therefore, once the kernel has been determined over a finite interval and has decayed sufficiently beyond that interval, the same kernel can be reused to propagate the dynamics to later times.

It is worth noticing that finite-memory NZ/TTM provides an explicit nonlocal
evolution rule based on a sliding window of past states.  By contrast, a purely
TL extrapolation requires either an explicit model
for $\mathcal{L}(t)$ at future times or a clear identification of a time
interval where the generator has become effectively time independent. In the present example the rapid decay of $\mathcal{K}(t)$ provides a direct and robust route to long-time propagation.

\subsection{Population-only memory kernels and Pauli master equations}
\label{subsec:population_kernel}

In many chemical-physics applications, one ultimately seeks an effective kinetic
description in terms of populations in a chosen basis (site
populations, or specific electronic states of a single molecular system).  
This description can be readily obtained 
starting from the full map $\Phi(t)$, and defining the
population projection superoperator
\begin{equation}
\mathcal{P}\rho=\sum_{m=1}^{8}\ket{m}\!\bra{m}\rho\ket{m}\!\bra{m},
\label{eq:pop_projector}
\end{equation}
and the population vector $p_m(t)=\mel{m}{\rho_S(t)}{m}$.
If $\rho(0)$ is diagonal, \textit{i.e.} contains only populations,
it is possible to generate the population map $M(t)$,
\begin{equation}
p(t) = M(t)\,p(0)
\label{eq:Mpop_def}
\end{equation}
where $M(t)_{i,j} = \Phi(t)_{ii,jj}$ is a $d_S\times d_S$ matrix, containing
the diagonal subset of the diagonal elements of the full map $\Phi(t)$.

Starting from Eq. \ref{eq:Mpop_def} one may construct a
time-local generator for populations, 
\begin{equation}
\dot p(t)=W(t)\,p(t), 
\end{equation}
where $W(t)$ can be interpreted as a matrix of \emph{instantaneous} population-transfer rates
and satisfies the equation
\begin{equation}
\dot M(t)  = W(t) M(t).
\label{eq:wtpop}
\end{equation}
In a regime where coherences have decayed and bath memory is short,
$W(t)$ becomes slowly varying; one may then approximate it by
a constant rate matrix $W$ over a suitable time window and propagate populations
with a Pauli master equation, \begin{equation}
\dot{p}_m(t)=\sum_{n=1}^{8} W_{mn}\,p_n(t).
\label{eq:Pauli}
\end{equation}

Defining the Nakajima--Zwanzig population kernel $\kappa(t)$
via the generalized master equation 
\begin{equation}
\dot p(t)=\int_0^t \kappa(t-s)\,p(s)\,ds.
\label{eq:NZ_pop}
\end{equation}
The elements of $\kappa(t)$ can be obtained using 
Eq. \ref{eq:kernel_rec} replacing $\Phi(t)$ by $M(t)$,
$W$ can be estimated
from the Laplace transform of the kernel
\begin{equation}
\label{eq:W_from_kappa}
W \approx \widehat \kappa(0^+) = \lim_{s\rightarrow 0^+} \int_0^{+\infty}
e^{-s t}\kappa(t)\,dt.
\end{equation}
If the process is significantly non-Markovian it is also reasonable
to time-average $W(t)$ over an interval where it is
approximately stationary\cite{LandiEtAl2018JCTC,VelardoEtAl2016JPCC,BorrelliPeluso2015JCTC} 

Alternatively we can define the best 
effective Pauli generator as a constrained least-squares solution of Eq. \ref{eq:wtpop}
\begin{equation}
W^\star
=
\underset{W\in\mathcal P}{\operatorname{argmin}}
\sum_{n=n_0}^{n_1}
\omega_n
\left\|
\dot M(t_n)-W M(t_n)
\right\|_F^2 ,
\label{eq:wfit}
\end{equation}
where \(\omega_n\) are optional weights and \(\|\cdot\|_F\) denotes the Frobenius norm. In practice, \(\dot M(t_n)\) is evaluated from the discrete population maps by finite differences. 
\(\mathcal P\) is the set of admissible Pauli rate matrices satisfying non-negative off-diagonal rates and population conservation, \(W_{jj}=-\sum_{i\neq j}W_{ij}\).

The lower limit \(t_{n_0}\) defines the initial time of the fitting window. Different choices of \(t_{n_0}\) generally lead to different fitted matrices, especially when the early dynamics contains non-Markovian transients or coherence--population coupling. In the present work, however, our goal is not to construct a family of local generators \(W(t)\), but rather to obtain a single global effective Pauli matrix. We therefore perform the fit over a fixed global time interval and interpret \(W^\star\) as the best constrained Pauli projection of the full population-map dynamics on that interval.
The complete $W^\star$ is provided in the SI file.

\begin{figure}[t]
  \centering
  \safeincludegraphics[width=0.9\linewidth]{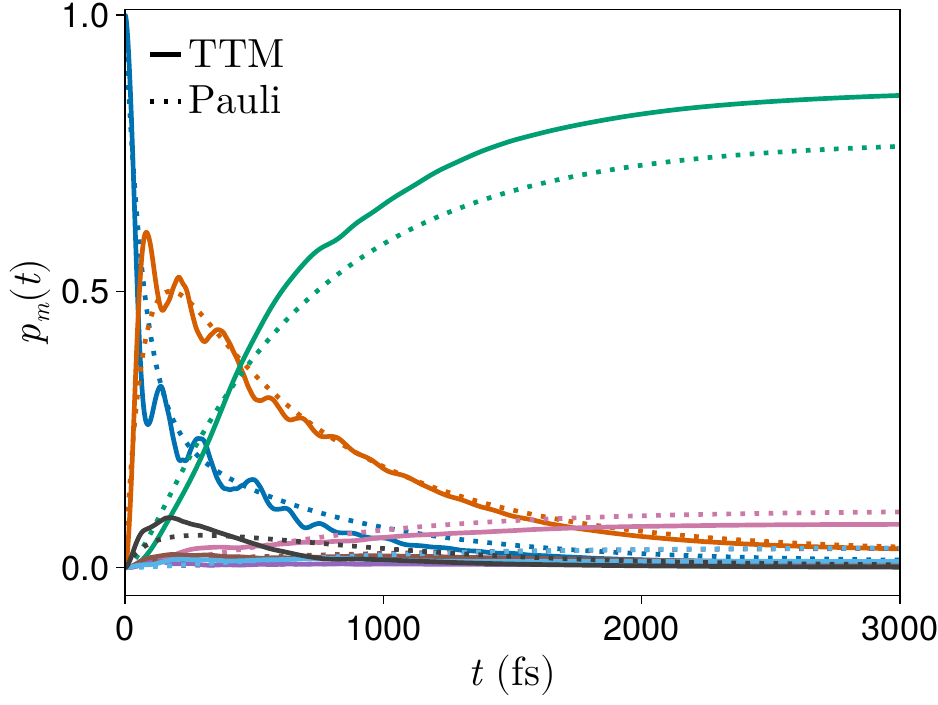}
  \caption{Population dynamics extrapolated up to 3~ps. The system 
  is initially excited at site 1. 
Solid lines: Finite-memory transfer-tensor propagation with memory cut at 600 fs;
Dotted lines: Pauli master equation~\eqref{eq:Pauli}.
}
  \label{fig:population_extrap}
\end{figure}

Fig.~\ref{fig:population_extrap} compares long-time population dynamics
obtained from (i) finite-memory transfer tensors and (ii) a time-local Pauli
equation with an effective constant $W$ estimated from Eq. \ref{eq:wfit}. 

The results highlight several important features of the exciton dynamics. 
As expected, the Pauli model fails to capture the early-time coherent dynamics,
which are strongly non-Markovian. 
By contrast, finite-memory TTM
propagation remains accurate throughout the initial transient and intermediate
regimes, where non-Markovian effects are still significant.
Moreover, the long-time 
values of the site populations
of the two models show  clear differences.

However, using the TTM method with a finite memory cutoff to extrapolate
long-time dynamics requires special care, as it guarantees neither CP nor TP 
unless memory effects are truly negligible. This occurs because
even small terms neglected in the NZ kernel can accumulate over long times, producing large errors. 
Conversely, 
if, after TTM extrapolation, the map remains CP, this indicates that the chosen memory window captures the essential memory time of the process. For the current FMO model, we have verified that the extrapolated map remains CP at least up to 3 ps.

Finally, it is worth noting
that in most cases the master-equation approach is not implemented via the
dynamical maps $\Phi(t)$, or $M(t)$, but instead by estimating approximate rates via
Fermi’s golden rule or related schemes, making such approaches extremely cheap
from a computational point of view.
Recently, Runeson \textit{et al.} \cite{RunesonEtAl2024PCCP} have
compared F\"orster, Redfield and surface hopping methods for the eight-site FMO model based on the same exciton Hamiltonian used in the present paper. However, because they have employed a Debye spectral density with a small reorganization energy, any comparison with the present fully quantum calculation is not possible. 
Previous works\cite{IshizakiFleming2009JCP,IshizakiFleming2009JCPa,WilkinsDattani2015JCTC} have shown that approximate kinetic theories for the FMO system are in reasonable agreement with the fully quantum calculation, however these comparisons were based on extremely simple spectral densities which 
might not be able at all to capture the non-Markovian behavior of the process.\cite{LorenzoniEtAl2025SA,ChinEtAl2013NP}

\section{Conclusions}
\label{sec:conclusions}

We have presented a practical route to compute the reduced dynamical
maps $\Phi(t)$ for complex system--bath models at finite temperature by
combining three ingredients:  (i) a Choi-matrix representation of the reduced
map, which encodes the complete reduced propagator as a $d_S^2\times d_S^2$
matrix;  (ii) TFD, which purifies the thermal bath and
converts finite-temperature reduced dynamics into a wavefunction propagation
problem in a doubled Hilbert space; and  (iii) TT propagation and
contractions, which make both the propagation and the required bath traces
feasible for high-dimensional structured environments.
For the entire procedure to be succesful, we have exploited the recently 
developed ID discretization technique that ensures an effective discretization 
of the bath correlation functions with a minimum number of modes in a given 
time range.\cite{TakahashiBorrelli2024JCP}

From a \emph{single} TT propagation of the thermofield initial state
$|\Omega\rangle\otimes|0\rangle$ under the Hamiltonian $H_\theta$, we obtain
the Choi matrix $J_{\Phi(t)}$ on a time grid by efficient TT contractions.  Once
stored, the map can be applied to arbitrary initial system density matrices at
negligible additional cost, enabling systematic post-processing of populations,
coherences and any other system dynamical variable without rerunning expensive
simulations.

Beyond propagation, the map-based viewpoint provides direct access to reduced
descriptions that are widely used in chemical physics.  From the same map data
one can in principle construct time-local generators and time-nonlocal memory kernels or transfer
tensors which can in turn be useful for extrapolation of long-time dynamics. 
A further reduction
to population dynamics yields population kernels and, when appropriate,
effective Pauli rate matrices that connect naturally to kinetic modeling.

Our application to exciton transport in the FMO complex illustrates these
capabilities.  Choi-spectrum and entropy diagnostics provide a compact
separation between an initial decoherence stage and a slower
population-relaxation stage, while the reconstructed memory kernels reveal a
finite bath memory time on the scale of a few hundred femtoseconds.  This
enables accurate long-time population extrapolation using finite-memory transfer
tensors and, at longer times, approximate time-local rate descriptions.
It should be noted that realistic spectral densities are important for a proper
understanding of non-Markovianity in molecular systems. While simplified models
(such as Ohmic, Debye, and Brownian oscillator forms) can be useful, they may
also bias intuition about the time scales needed for reliable long-time
extrapolation. Using more realistic spectral densities typically reveals that
longer dynamical information can be relevant.
 
Finally, the framework readily accommodates ensemble effects such as static
disorder by introducing auxiliary zero-frequency modes, allowing
disorder-averaged maps to be obtained without explicit sampling over many
disorder realizations.  Taken together, these elements make the TFD--Choi--TT
approach a flexible tool for building reduced propagators and kinetic models for
complex molecular systems with structured, finite-temperature environments.

The limits of the above computational framework are mostly in the numerical capabilities of obtaining
accurate dynamics from the time evolution of the TT. 
Although this can be achieved only for relatively short times, the rapid decay
of the memory in the NZ kernel makes the time-nonlocal approach amenable for an
approximate yet meaningful extrapolation of the dynamics to long times. 
Recent advances in the determination of TT time evolution can also be of help
and are currently being investigated.\cite{NuominEtAl2022PRA}

Finally, we mention that since the determination of the
local generator can exhibit numerical stability issues,
alternative approaches will be investigated in the future 
which bypass the matrix inversion step.\cite{NestmannEtAl2021PRX}

\section*{Supplementary Material}
The supplementary material provides additional details on the derivation of the TFD Hamiltonian operator; the figures showing the 
convergence of the the ID procedure; the parameters used in the TT-TDVP time evolution; a study of the TP error and the Pauli rate matrix obtained as described in the text.

\appendix

\section{Tracing the bath using TT format}
\label{sec:tt_trace}

For completeness, here we report the algorithm used to compute 
partial traces in TT format. 
Assume we have the thermal state $|\varphi\rangle$ in TT format 
as described in Section \ref{sec:tt_rep}.
We can explicitly write the partial trace over the environment states in 
Eq. \eqref{eq:choitfd} as 
\begin{equation}
    J_\Phi(i_a,i_s,j_a,j_s) 
=\!\!\!\! \sum_{k_1,\dots,k_{N_B}}\!\!\!\!
\varphi_{i_a,i_s,k_1,\dots,k_{N_B}}\;\overline{\varphi}_{j_a,j_s,k_1,\dots,k_{N_B}}
\end{equation}
where the bar denotes complex conjugation.

To exploit the TT structure of Eq. \ref{eq:phitt} in the bath contraction
we start from the last bath site, and construct the matrix
\begin{equation}
E^{(N_B+2)}_{\alpha_{N_B+1},\alpha'_{N_B+1}} =
 \sum_{k_{N_B}}
   G^{(N_B+2)}_{\alpha_{N_B+1},k_{N_B},1}\;
   \overline{G}^{(N_B+2)}_{\alpha'_{N_B+1},k_{N_B},1}.
\end{equation}
After this initial matrix is computed we sweep from right to left for $t=N_B-1,\dots,1$ and compute the matrices
\begin{multline}
E^{(t+2)}_{\alpha_{t+1},\alpha'_{t+1}}
= \sum_{\alpha_{t+2},\alpha'_{t+2}}
   E^{(t+3)}_{\alpha_{t+2},\alpha'_{t+2}} \\
   \times \left(\sum_{k_t}
   G^{(t+2)}_{\alpha_{t+1},k_t,\alpha_{t+2}}\;
   \overline{G}^{(t+2)}_{\alpha'_{t+1},k_t,\alpha'_{t+2}} \right)
\end{multline}
At the end of the right-left sweep we have contracted all the cores to the matrix $E^{(3)}_{\alpha_2,\alpha'_2}$.
Finally, we have to contract over the bond dimensions of the ancilla and of the system.
To this end we first create the tensor
\begin{equation}
M^{(2)}_{\alpha_1,\alpha'_1;\; i_s,j_s;\; \alpha_2,\alpha'_2}
= G^{(2)}_{\alpha_1, i_s, \alpha_2}\;
  \overline{G}^{(2)}_{\alpha'_1, j_s, \alpha'_2}.
\end{equation}
then we  absorb environment $E^{(3)}$ tensor:
\begin{equation}
\widetilde M^{(2)}_{\alpha_1,\alpha'_1;\; i_s,j_s} 
= \sum_{\alpha_2,\alpha'_2}
  M^{(2)}_{\alpha_1,\alpha'_1;\; i_s,j_s;\; \alpha_2,\alpha'_2}\;
  E^{(3)}_{\alpha_2,\alpha'_2}.
\end{equation}
As a last step we contract $\widetilde M^{(2)}$ with the 
ancilla cores:
\begin{equation}
M^{(1)}_{i_a,j_a;\; \alpha_1,\alpha'_1}
= G^{(1)}_{1,i_a,\alpha_1}\;
  \overline{G}^{(1)}_{1,j_a,\alpha'_1}.
\end{equation}
We then obtain the Choi matrix as:
\begin{equation}
J_\Phi(i_a,i_s;j_a,j_s)
= \sum_{\alpha_1,\alpha'_1}
  M^{(1)}_{i_a,j_a;\,\alpha_1,\alpha'_1}\;
  \widetilde M^{(2)}_{\alpha_1,\alpha'_1;\, i_s,j_s}.
\end{equation}
We note that for $J_\Phi$ to be a matrix each pair of indices $(i_a,i_s)$ and $(j_a,j_s)$ is flattened to a single index.

\section*{Acknowledgements}
% Acknowledge funding, discussions, or support here.
This work was partially supported by the Spoke 7 “Materials and Molecular
Sciences” of ICSC – Centro Nazionale di Ricerca in High-Performance Computing,
Big Data, and Quantum Computing, funded by European Union—NextGenerationEU. 

R.B. acknowledges the research project “nuovi Concetti, mAteriali e tecnologie
per l’iNtegrazione del fotoVoltAico negli edifici in uno scenario di generazione
diffuSa” (CANVAS), funded by the Italian Ministry of the Environment and the
Energy Security, through the Research Fund for the Italian Electrical System
(type-A call, published on G.U.R.I. n. 192 on 18-08-2022).

\section*{Author contributions}
R.B. conceived the project and developed the theoretical framework. 
R.B. and H.T. implemented the software and wrote the manuscript.

\section*{Competing interests}
The authors declare no competing interests.

\section*{Data availability} 
All data supporting the findings of this study are available at the address:
\href{https://doi.org/10.5281/zenodo.19855477}{10.5281/zenodo.19855477}

% ====================
% REFERENCES
% ====================
 \bibliography{cit}

\end{document}